\documentclass[a4paper]{article}  
\pdfoutput=1
\usepackage{graphicx}
\usepackage{amssymb}
\usepackage{lineno}
\usepackage{authblk}
\usepackage{hyperref}

\begin{document}

\title{Long-term monitoring of the ANTARES optical module efficiencies using $^{40}\mathrm{K}$ decays in sea water}

\author[a]{A.~Albert}
\author[b]{M.~Andr\'e}
\author[c]{M.~Anghinolfi}
\author[d]{G.~Anton}
\author[e]{M.~Ardid}
\author[f]{J.-J.~Aubert}
\author[g]{J.~Aublin}
\author[g]{T.~Avgitas}
\author[g]{B.~Baret}
\author[h]{J.~Barrios-Mart\'{\i}}
% ...

\affil[a]{\scriptsize{Universit\'e de Strasbourg, CNRS,  IPHC UMR 7178, F-67000 Strasbourg, France}}
\affil[b]{\scriptsize{Technical University of Catalonia, Laboratory of Applied Bioacoustics, Rambla Exposici\'o, 08800 Vilanova i la Geltr\'u, Barcelona, Spain}}
\affil[c]{\scriptsize{INFN - Sezione di Genova, Via Dodecaneso 33, 16146 Genova, Italy}}
\affil[d]{\scriptsize{Friedrich-Alexander-Universit\"at Erlangen-N\"urnberg, Erlangen Centre for Astroparticle Physics, Erwin-Rommel-Str. 1, 91058 Erlangen, Germany}}
\affil[e]{\scriptsize{Institut d'Investigaci\'o per a la Gesti\'o Integrada de les Zones Costaneres (IGIC) - Universitat Polit\`ecnica de Val\`encia. C/  Paranimf 1, 46730 Gandia, Spain}}
\affil[f]{\scriptsize{Aix Marseille Univ, CNRS/IN2P3, CPPM, Marseille, France}}
\affil[g]{\scriptsize{APC, Univ Paris Diderot, CNRS/IN2P3, CEA/Irfu, Obs de Paris, Sorbonne Paris Cit\'e, France}}
\affil[h]{\scriptsize{IFIC - Instituto de F\'isica Corpuscular (CSIC - Universitat de Val\`encia) c/ Catedr\'atico Jos\'e Beltr\'an, 2 E-46980 Paterna, Valencia, Spain}}
% ...

\date{}
\maketitle
\thanks{\href{mailto:salvadori@cppm.in2p3.fr}{salvadori@cppm.in2p3.fr}}

\newpage
\begin{abstract}
Cherenkov light induced by radioactive decay products is one of the major sources of background light for deep-sea neutrino telescopes such as ANTARES. These decays are at the same time a powerful calibration source. Using data collected by the ANTARES neutrino telescope from mid 2008 to 2017, the time evolution of the photon detection efficiency of optical modules is studied. A modest loss of only 20\% in 9 years is observed. The relative time calibration between adjacent modules is derived as well.
\end{abstract}

\section{Introduction}
\label{intro}
The ANTARES neutrino telescope~\cite{Ant} aims at the exploration of the high-energy Universe by using neutrinos as cosmic probes. Three main search methods are exploited. The first one refers to searches for steady neutrino sources, through the identification of an excess of events from a given sky direction~\cite{ps}. The second one searches for an excess of high-energy cosmic neutrinos over the background of atmospheric events~\cite{diffuse}. The third method looks for temporal and/or spatial correlations with transient events observed through other probes. An example of this last method is offered by the searches for neutrinos in concomitance with GW170817~\cite{multim}. In addition, the ANTARES detector is used for particle physics studies, such as neutrino oscillations~\cite{osci}, search for magnetic monopoles~\cite{mm}, and indirect searches for dark matter~\cite{sun,gc}.

For all these studies, a continuous monitoring of the positioning of all parts of the detector is needed in order to preserve the precision on the reconstruction of neutrino-induced events. Moreover, it is necessary to check the stability of the optical sensors as a function of time, in order to guarantee an equable estimate of the energy released by charged particles. Studies on the performance as well as on the stability of optical sensors of this kind have been discussed also by other Collaborations (e.g.~\cite{ICOM,SKOM}). The originality of the present paper relies on the use of a natural calibration tool offered by the presence of radioactive elements dissolved in sea water.

\begin{figure}[h!]
\begin{minipage}[b]{1.\linewidth}
\centering
\includegraphics[width=0.5\linewidth]{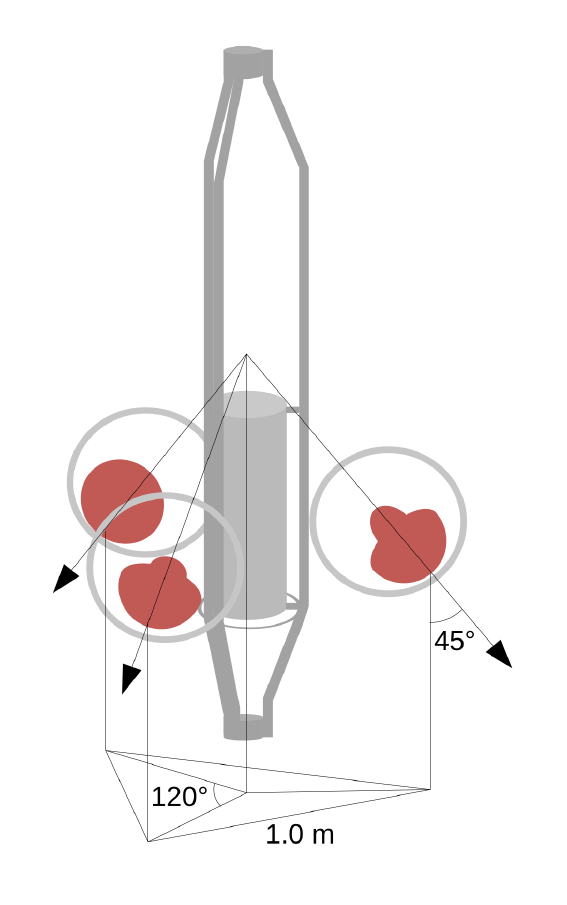}
\caption{Schematic representation of an ANTARES storey. The spheres stand for the OMs, which contain one PMT each, facing 45$^{\circ}$ downwards.}
\label{fig1}
\end{minipage}
\end{figure}

The ANTARES neutrino telescope is located in the Mediterranean Sea, 40\,km off the coast of Toulon, France, at a mooring depth of about 2475\,m. The detector was completed in 2008. It is composed of 12 detection lines, each one equipped with 25 storeys of 3 optical modules (OMs), except line 12 with only 20 storeys of OMs, for a total of 885 OMs. The horizontal spacing among the lines is $\sim$60\,m, while the vertical spacing between the storeys is 14.5\,m. Each OM hosts a photomultiplier tube (PMT), whose axis points 45$^{\circ}$ downwards (see Figure~\ref{fig1}). The PMTs are 10-inch tubes from Hamamatsu~\cite{Hamamatsu}. The relative position of each OM in the detector is monitored in real time by a dedicated positioning system~\cite{posi}. All signals from the PMTs that pass a threshold of 0.3 single photoelectrons (hits) are digitised and sent to the shore station~\cite{Elec,DA}.

\newpage

The decay products of radioactive elements dissolved in sea water constitute the principal source of background light for deep-sea neutrino telescopes. Among these isotopes, potassium-40 ($^{40}\mathrm{K}$) is the most abundant. Other radioactive decays (mainly from the U/Th chain) induce Cherenkov photons on the permille level compared to $^{40}\mathrm{K}$ decays and can be neglected. This process constitutes an important calibration tool as well. If a $^{40}\mathrm{K}$ nucleus decays near a storey, the resulting Cherenkov light can be recorded by two OMs almost simultaneously. Such coincidences are used to derive the relative photon detection efficiencies~\cite{K40} and for time calibration between OMs in the storey.\\
\indent The document is organised as follows: in Section~\ref{sec:1} the $^{40}\mathrm{K}$ decay rate in sea water is computed and its stability is discussed; the method for the OM photon detection efficiency computation using $^{40}\mathrm{K}$ decays is described in Section~\ref{sec:2}, together with the data sample employed, while the results of the analysis are presented in Section~\ref{sec:3}. A brief explanation on the use of the $^{40}\mathrm{K}$ decay measurements for the time calibration of the detector is given in Section~\ref{sec:4}, while in Section~\ref{sec:5} conclusions are presented.

\section{$^{40}\mathrm{K}$ decay rate at the ANTARES site}
\label{sec:1}
The main decay channels of $^{40}\mathrm{K}$ are:
\[^{40}\textrm{K} \rightarrow \hspace*{0.1cm} ^{40}\hspace{-0.04cm}\textrm{Ca} + e^- + \overline{\nu}_e \hspace*{2.cm} (89.3\%)\]
\[^{40}\textrm{K} + e^- \rightarrow \hspace*{0.1cm}  ^{40}\hspace{-0.04cm}\textrm{Ar}^* + \nu_e \hspace*{1.9cm} (10.7\%)\]
\vspace*{-0.2cm}
\hspace*{4.7cm} $\hookrightarrow  \hspace*{0.1cm}  ^{40}\hspace{-0.04cm}\textrm{Ar} + \gamma$

\vspace*{0.3cm}
The electron produced in the $\beta$-decay channel, with an energy up to 1.3\,MeV, leads to the production of Cherenkov light when traveling in water. In the electron capture channel, fast electrons with subsequent Cherenkov light emission are produced by Compton scattering of the 1.46\,MeV photon, released by the excited Ar nuclei.

The detection rate, $R_s$, of Cherenkov photons from products of $^{40}\mathrm{K}$ decays on an ANTARES optical module can be factorised as:

\begin{equation}
R_s = B_q \cdot V_s,
\end{equation}

\noindent with $B_q$ the $^{40}\mathrm{K}$ decay rate per unit volume and $V_s$ the effective volume around a single OM for detecting a single photon from a $^{40}\mathrm{K}$ decay. As there is no preferred decay direction, $V_s$ can be calculated in a semi-analytical way, by integrating over the decay positions around the OM. One obtains, for a given wavelength $\lambda$, a factorisation into a wavelength-dependent effective area $A(\lambda)$ and the corresponding photon absorption length $L_{abs}(\lambda)$ of sea water. $V_s$ can then be written as:

\begin{equation}
V_s = A_s \cdot L_{abs}^0,
\end{equation}

\noindent with $A_s = \int A(\lambda)\Phi(\lambda)L_{abs}(\lambda) d\lambda/\Phi_{tot}L^{0}_{abs}$, with $\Phi(\lambda)$ containing the --- arbitrarily normalised --- $\lambda$ dependence of the Cherenkov photon flux (close to $1/\lambda^2$), $\Phi_{tot}$ its integral over a given wavelength range and $L_{abs}^0$ the absorption length at some reference wavelength. $A(\lambda)$ is determined by simulating an isotropic photon flux around an OM and depends on the OM properties such as the size of its photocathode, its quantum efficiency, and its angular acceptance. The PMT quantum efficiency depends on the photon wavelength and it is deduced from the product sheet as reported in~\cite{Hamamatsu}. The absorption length depends on the photon wavelength as well as parameterized in~\cite{AbsL}.\\
\indent The integration to evaluate $A_s$ is carried out in the wavelength range 300-600\,nm, yielding $A_s = 420 \pm 50$\,cm$^2$. The total uncertainty is dominated by uncertainties on the mentioned PMT properties (in particular, photocathode size and angular acceptance).

If a $^{40}\mathrm{K}$ nucleus decays near a storey, the associated Cherenkov light can be recorded by two OMs almost simultaneously: this kind of signal is referred to as \textit{genuine} coincidence. Its rate can be written as:

\begin{equation}
R_c = B_q \cdot V_c,
\label{eq:Eq3}
\end{equation}

\noindent where $V_c$ is an effective volume around a pair of OMs for the detection of a coincident signal from a single $^{40}\mathrm{K}$ decay. The value of $V_c$ is derived from Geant-3 simulations, modelling $^{40}\mathrm{K}$ decays around a pair of OMs and propagating the decay products through the sea water with a full tracking of electrons, including multiple scattering and velocity dependent emission of Cherenkov light.
These simulations yield $V_c~=~1100~\pm~370$\,cm$^3$. The relative uncertainty of $V_c$ is significantly larger compared to $A_s$ for two reasons. Firstly, uncertainties from two OMs contribute to $V_c$ whereas only one OM contributes to $A_s$. Secondly, as the axes of adjacent OMs are separated by 75.5$^{\circ}$, the uncertainty on the OM angular acceptance has a much larger influence for $V_c$. The calculation of $A_s$, on the other hand, only depends on its integral value.

Whereas $^{40}\mathrm{K}$ decays can contribute to the rate of a single OM, $R_s$, up to a distance of $\sim L_{abs}^0$, contributions to the rate of genuine coincidences, $R_c$, are confined to a small volume in the vicinity of the storey, with 90\% occuring within 3\,m. At such a distance, the effect of absorption is below 5\% and the associated uncertainties become negligible. The dominating error for $R_c$ originates from the uncertainties on the OM properties. Therefore the measurement of the coincidence signals can be directly related to the OM efficiencies. Due to the large uncertainty of $V_c$, only relative efficiencies are determined in the following.

Both  $R_s$ and $R_c$ depend on $B_q$, with:

\begin{equation}
B_q = r_s \cdot r_K \cdot r_I \cdot \rho \cdot \frac{\ln2}{\tau_{1/2}} \cdot \frac{N_A}{A},
\end{equation}

\noindent where $N_A$ is the Avogadro number, $A=39.96$~\cite{AtMassK40} and $\tau_{1/2}=1.25\times10^9$\,years~\cite{HalfT} are the atomic mass and lifetime of $^{40}\mathrm{K}$, and $\rho = 1.038$\,g/cm$^3$ is the density of deep-sea water at the ANTARES site, which is derived from in situ measurements of pressure, temperature and salinity, $r_s$. The parameter $r_K$ is the potassium fraction in Mediterranean Sea salt and $r_I$ is the isotope fraction of $^{40}\mathrm{K}$. The last three quantities can be considered stable over the lifetime of ANTARES with variations smaller than 1\%. The salinity is monitored directly with the ANTARES instruments and found to be $r_s = 3.844\%$, while $r_K = 1.11\%$ from~\cite{KSeaWat} and $r_I~=~1.17~\cdot~10^{-4}$ from~\cite{K40inK}.\\This yields $B_q~=~13700~\pm~200$\,s$^{-1}$m$^{-3}$.

With this value, $R_s = 35 \pm 8$\,kHz and $R_c = 15 \pm 5$\,Hz are the single and genuine coincidence rates obtained from simulations. The observed single photon rates in ANTARES OMs are about 55\,kHz, due to additional light from bioluminescence that cannot be filtered out at the single photon level. This phenomenon can result in variations of the single photon rates.  Occasionally, this can produce short bursts (lasting few seconds) with single photon rates up to several orders of magnitude higher than due to $^{40}\mathrm{K}$ decays~\cite{BioBursts}. On the contrary, no variability on similar time scales is observed for the genuine coincidence rates. This seems in agreement with the fact that bioluminescence is emitted from the reaction of luciferase enzyme occurring at level of single photon emission only; in any case, bioluminescence does not produce a significant amount of correlated photons on the nanosecond level in our OMs.\\
\indent Coincidence signals from low-energy muons do not exceed 1\% compared to the $^{40}\mathrm{K}$.

\section{Detection efficiency calibration using $^{40}\mathrm{K}$}
\label{sec:2}
Data collected from mid 2008 to December 2017 have been analysed in this work. The $\mathrm{^{40}K}$ trigger selects coincident photons in adjacent OMs if they are detected within a narrow time window of 50\,ns.

From the completion of the detector in May 2008 until November 2009, dedicated runs with $\mathrm{^{40}K}$ triggers have been taken once per week, with a down-scaling factor of 4 in order not to saturate the readout data acquisition system. Runs of the same month have been merged together. In November 2009 the $\mathrm{^{40}K}$ trigger has been integrated into the
standard data-taking setup with a down-scaling factor of 200. Now, each month of data taking has been divided into five periods
each providing a data sample of similar statistics as before.

\begin{figure}[h!]
\centering
\begin{minipage}[b]{1.\linewidth}
\includegraphics[width=\linewidth]{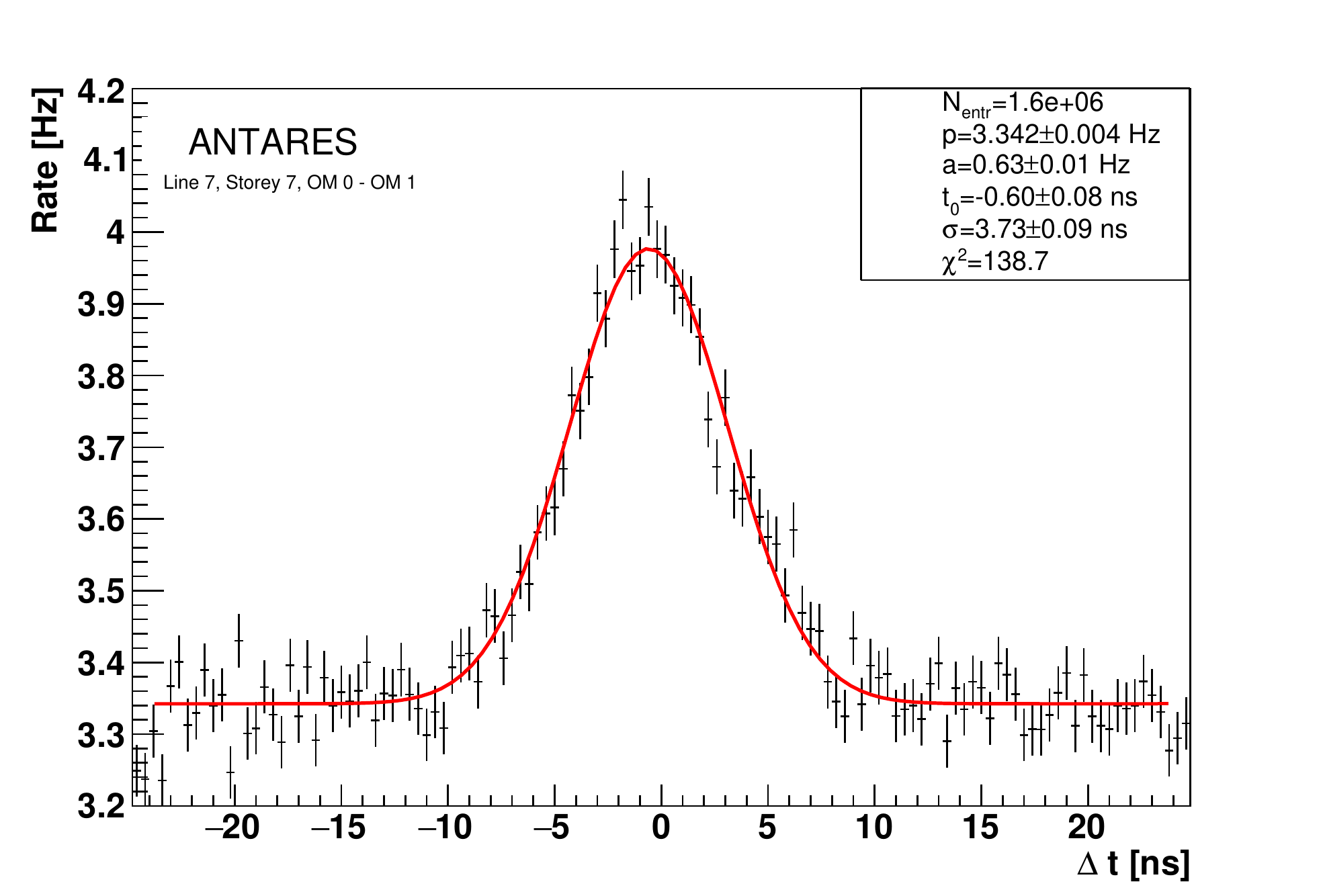}
\caption{Example of the detected hit time differences, $\Delta t$, between two adjacent OMs. The fitted parameters are listed as well (see Equation~\ref{eq:fit} for details). The plot refers to one pair of OMs in the same storey  (Line 7, Storey 7, OM 0 - OM 1) for one of the periods considered in the analysis.}
\label{fig2}
\end{minipage}
\end{figure}

Both the genuine coincidences induced by the same $^{40}\mathrm{K}$ decay process and random background coincidences, which are due to distinct $^{40}\mathrm{K}$ decays as well as to bioluminescence effects, contribute to hit pairs on adjacent OMs with small time differences $\Delta t$. Figure~\ref{fig2} shows an example of the distribution of the time difference between hits on two adjacent OMs. A Gaussian peak from genuine coincidences is clearly visible on top of a flat pedestal from uncorrelated coincidences.

The distribution of the coincidence signals is fitted with a Gaussian distribution added to a constant:

\begin{equation}\label{eq:fit}
f(t) = p + a\cdot\exp({-\frac{(t-t_0)^2}{2\sigma^2}}),
\end{equation}

\noindent where $p$ is the baseline, $a$ the amplitude of the Gaussian peak due to genuine coincidences, $\sigma$ is the peak width and $t_0$ the residual time offset between the hits on adjacent OMs. A value of $\sigma \sim 4$\,ns is expected, mainly due to the distance between the OMs and the spatial distribution of detected $\mathrm{^{40}K}$ coincidences around the storey.
A simple estimation can be obtained by considering the difference in the distance traveled by two photons emitted at the same position and detected by two different PMTs. In ANTARES the distance between the centers of photocathodes of two PMTs of the same storey is $l = 1.0$\,m (see Figure~\ref{fig1}), while the photocathode diameter $d$ is about 25\,cm. Therefore, neglecting light scattering, the maximum traveled path difference for two photons is $l + d = 1.25$\,m. Given the Cherenkov light velocity of $v_g = 0.217$\,m/ns, the corresponding time difference is about
5.8\,ns. By averaging over all the $\mathrm{^{40}K}$ disintegration positions which yield a genuine coincidence, a value of $\sigma=4$\,ns is obtained, compatible with Figure~\ref{fig2}.

For perfectly calibrated OMs, $t_0$ would be expected at 0\,ns. Deviations from the expected value of $t_0$ are mainly due to different PMT transit times. This makes the $\mathrm{^{40}K}$ method an integral part of the time calibration procedure, as will be discussed in more detail in Section~\ref{sec:4}. 

It has been verified that the parameter $p$ is slightly positively correlated with the parameter $a$, but not with the product $a\cdot \sigma$. In addition, both $a\cdot \sigma$ and $p$ parameters are not correlated with $t_0$.

The fit parameters can be used to estimate the number of events corresponding to the peak area, as:

\begin{equation}\label{eq:N}
R = \frac{a\cdot\sigma\cdot\sqrt{2\pi}}{\Delta \tau},
\end{equation}

\noindent where $\Delta \tau = 0.4$\,ns is the bin width used for the histogram.

For each storey three coincidence rates are measured ($R_{01}$, $R_{12}$ and $R_{20}$). These quantities are directly related to the photon detection efficiency of the three OMs ($\epsilon_0$, $\epsilon_1$ and $\epsilon_2$):

\begin{equation}
R_{ij}=R^*_c \cdot \epsilon_i \cdot \epsilon_j,
\label{eq:Rs}
\end{equation}

\noindent where $R^*_c$ is the rate for two nominal OMs with efficiencies equal to 1. A value of $R^*_c=15$\,Hz is used. This value is obtained as an average detector coincidence rate at the beginning of the analysed data set, and coincides with what is found in simulation. Solving the system of three equations, the corresponding efficiencies are derived:		

\begin{equation}\label{eq:Sens}
\epsilon_i = \sqrt{\frac{1}{R^*_c}\frac{R_{ij}\cdot R_{ki}}{R_{jk}}}.
\end{equation}

When an OM in a storey is not working for a given period, only one coincidence rate is measured, which is not sufficient to determine the two efficiencies. In this case, equal efficiencies for the two working OMs are assumed, namely:

\begin{equation}\label{eq:Sequal}
\epsilon_i = \epsilon_j = \sqrt{\frac{R_{ij}}{R^*_c}}.
\end{equation}

If two OMs in a storey are inactive no coincidences can be measured. In this case the average efficiency value of the line hosting that particular storey is assigned to the working OM for this period.

All coincidence histograms for all periods have been fitted according to Equation~\ref{eq:fit}. Data quality selection criteria have been applied, to ensure stable and reliable input for the subsequent efficiency calculation. A cut on the number of entries on each histogram, $N_{entr}$, excludes from the analysis all those cases for which the fit fails due to lack of statistics.  This includes the cases, for instance, in which one of the two OMs is only partially active. Taking into account that there are four fitted parameters and 120 bins in each histogram, a $\chi^2$ of 116 is expected for a good fit. Histograms with $\chi^2\geqslant200$ are excluded. Additional cuts on the fitted amplitude value and its uncertainty, $\Delta a$, have been applied to ensure a clear signal above the background. Furthermore, expected values of the Gaussian mean and width are known, thus cuts on these parameters have been applied. All the selection criteria applied are reported in Table~\ref{tab:Cuts}.

\begin{table}[h!]
\centering
\begin{tabular}{c}
\hline
Accepted Values \\
\hline
$N_{entr} > 2000$ \\
$\chi^2 < 200$ \\
$a > 0.1$\,Hz \\
$\Delta a/a < 0.1$ \\
$1.5$\,ns $< \sigma < 6.5$\,ns \\
$|t_0| < 20.0$\,ns \\
\hline
\end{tabular}
\caption{Values of data selection criteria applied in the analysis. The parameter nomenclature corresponds to the one used in Equation~\ref{eq:fit}.}
\label{tab:Cuts}
\end{table}

After applying these cuts, the efficiency can be determined on average for 77\% of those OMs which are active in a given period. 
The efficiency of the remaining working OMs is derived from adjacent OMs or from the average of the corresponding detection line as described above.

\section{Results}
\label{sec:3}
\begin{figure}[h!]
\centering
\begin{minipage}[b]{1.\linewidth}
\includegraphics[width=\linewidth]
{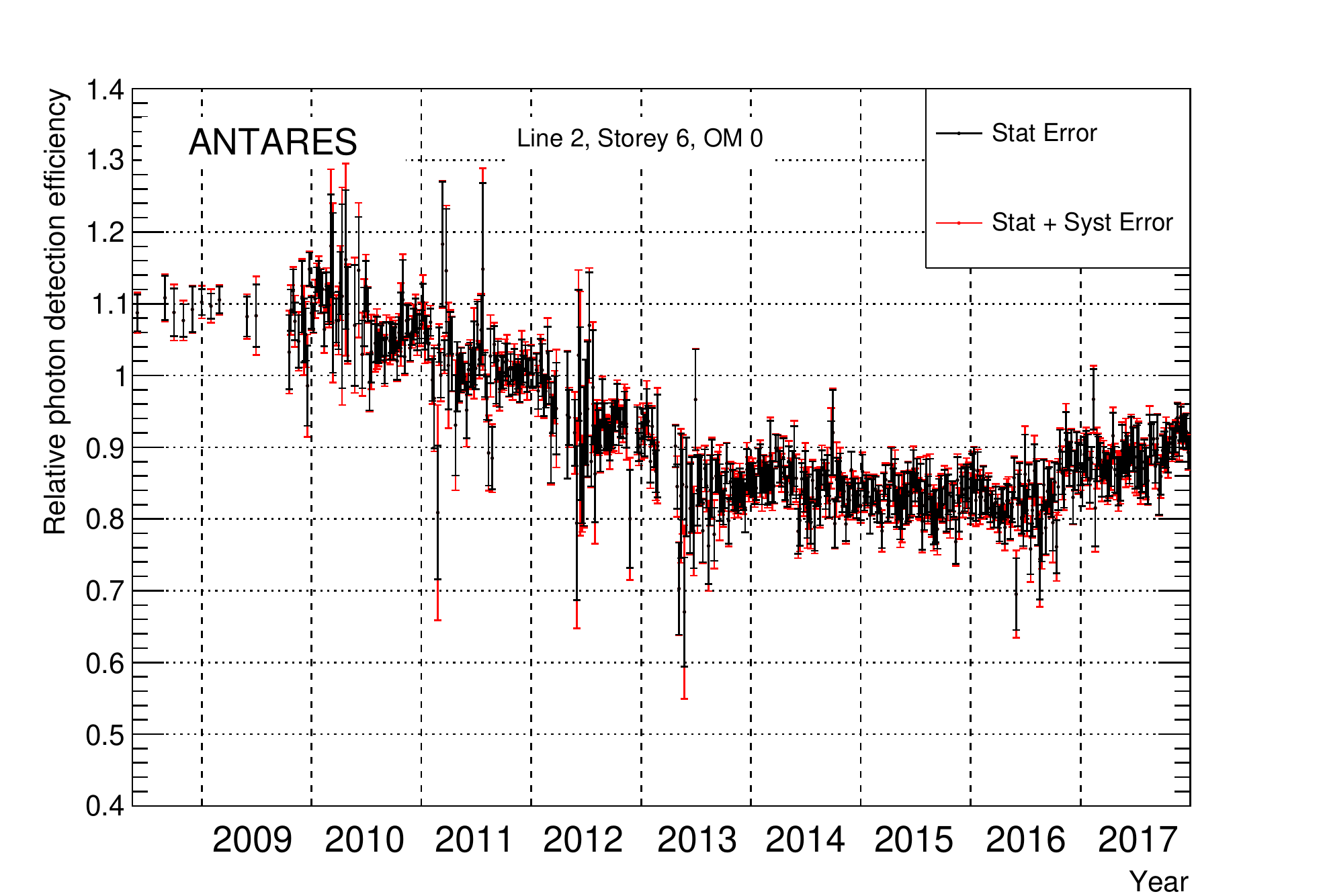}
\caption{Photon detection efficiency as a function of time for one ANTARES OM (Line 2, Storey 6, OM 0). Error bars correspond to statistical (black) and statistical plus systematic (red) contributions.}
\label{fig3}
\end{minipage}
\end{figure}

Histograms passing the quality criteria are then used to compute the OM photon detection efficiency, as described in Section~\ref{sec:2}. Figure~\ref{fig3} shows the photon detection efficiency as a function of time for one OM as an example. The uncertainty on the resulting photon detection efficiency of each OM has been calculated from the uncertainty on the fitted parameters and it is found to be around 3\% for all analysed periods. The total uncertainty is dominated by the statistical error. The systematic uncertainties (discussed later in this section) contribute on average only about 20\% of the total uncertainty. 

\newpage

\begin{figure}[h!]
\centering
\begin{minipage}[b]{1.\linewidth}
\includegraphics[width=\linewidth]
{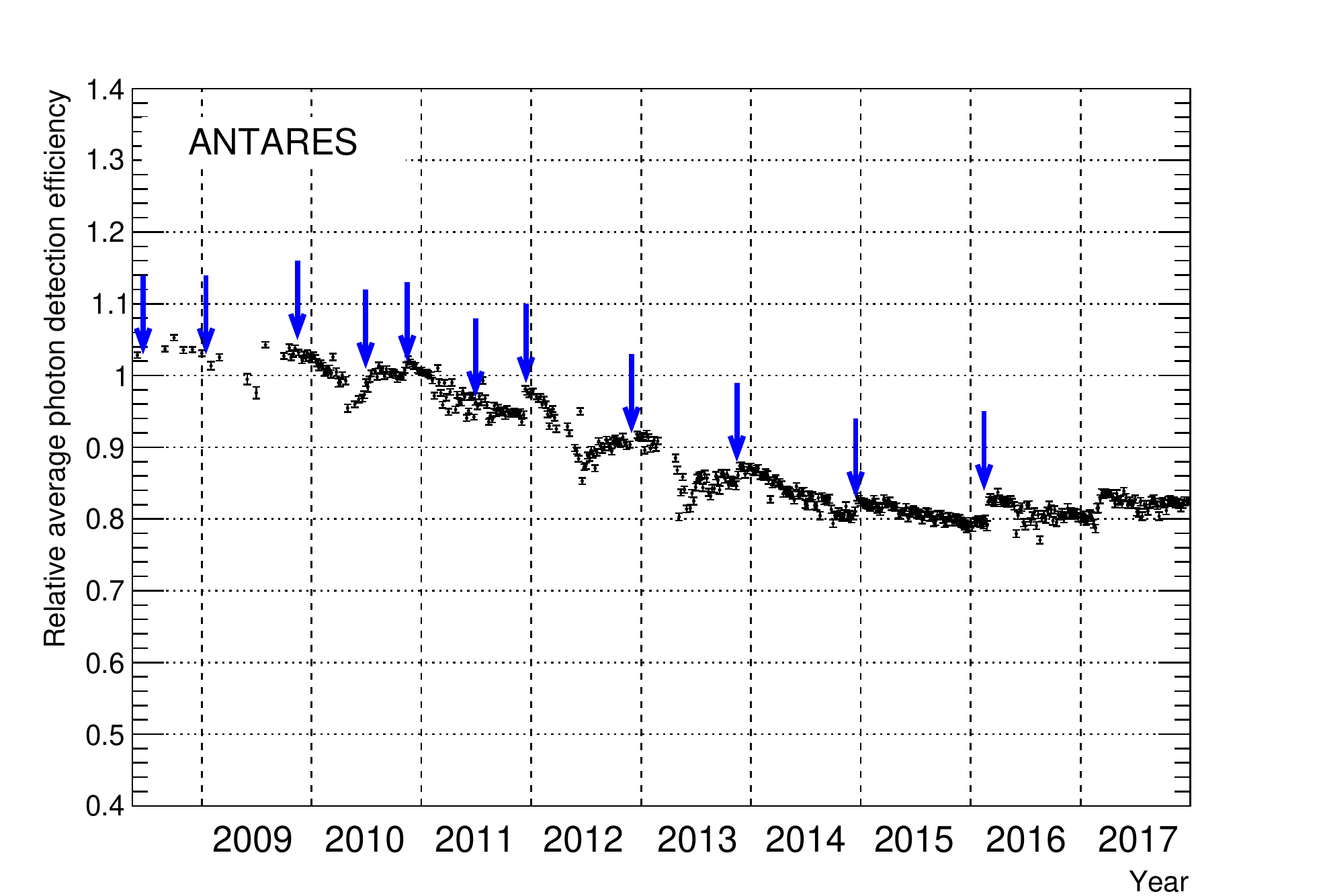}
\caption{Relative OM efficiency averaged over the whole detector as a function of time. The blue arrows indicate the periods in which high voltage tuning of the PMTs has been performed, while error bars indicate the statistical error $\sigma_{\rm mean}$ on the mean efficiency.}
\label{fig4}
\end{minipage}
\end{figure}

In order to monitor the status of the whole detector over several years, the average photon detection efficiency $\overline \epsilon$ as a function of time has been determined. For each period, the average $\overline \epsilon$ is computed over all OMs with non-zero efficiency. The result is shown in Figure~\ref{fig4}. It can be seen that, after a decrease over several years, the OM photon detection efficiency has finally stabilised over the last years. An overall modest detection efficiency loss of 20\% is observed over the whole analysed time period. In 2010, 2012 and 2013 a particular pattern is observed. The average OM photon detection efficiency drops by 5-10\% in spring and partly recovers in the second half of the year. This might be related to the formation of dense deep-sea water through a process known as "open-sea convection"~\cite{Tamburini}. As a consequence of such an exchange of deep sea water, sedimentation as well as biofouling processes might impact the OM photon detection efficiencies in these periods.

The distribution of the detected charge is regularly monitored for all PMTs and if either a significant broadening or a shift from the nominal position of the peak due to a single photoelectron is noticed a recalibration is performed by means of high voltage tuning (HVT). The HVT procedure adjusts the effective gain of individual PMTs to the nominal one; the procedure thus prevents detection effeciency losses due to a low gain and bias on the trigger logics. The effects of the HVT procedure, which is performed once or twice per year, are evident on the time dependence of  $\overline \epsilon$ shown in Figure~\ref{fig4}. From 2014 on, the observed pattern is clearly correlated to the HVT campaigns and the modest annual photon detection efficiency loss is fully recovered every year.

\newpage

\begin{figure}[h!]
\centering
\begin{minipage}[b]{1.\linewidth}
\includegraphics[width=\linewidth]
{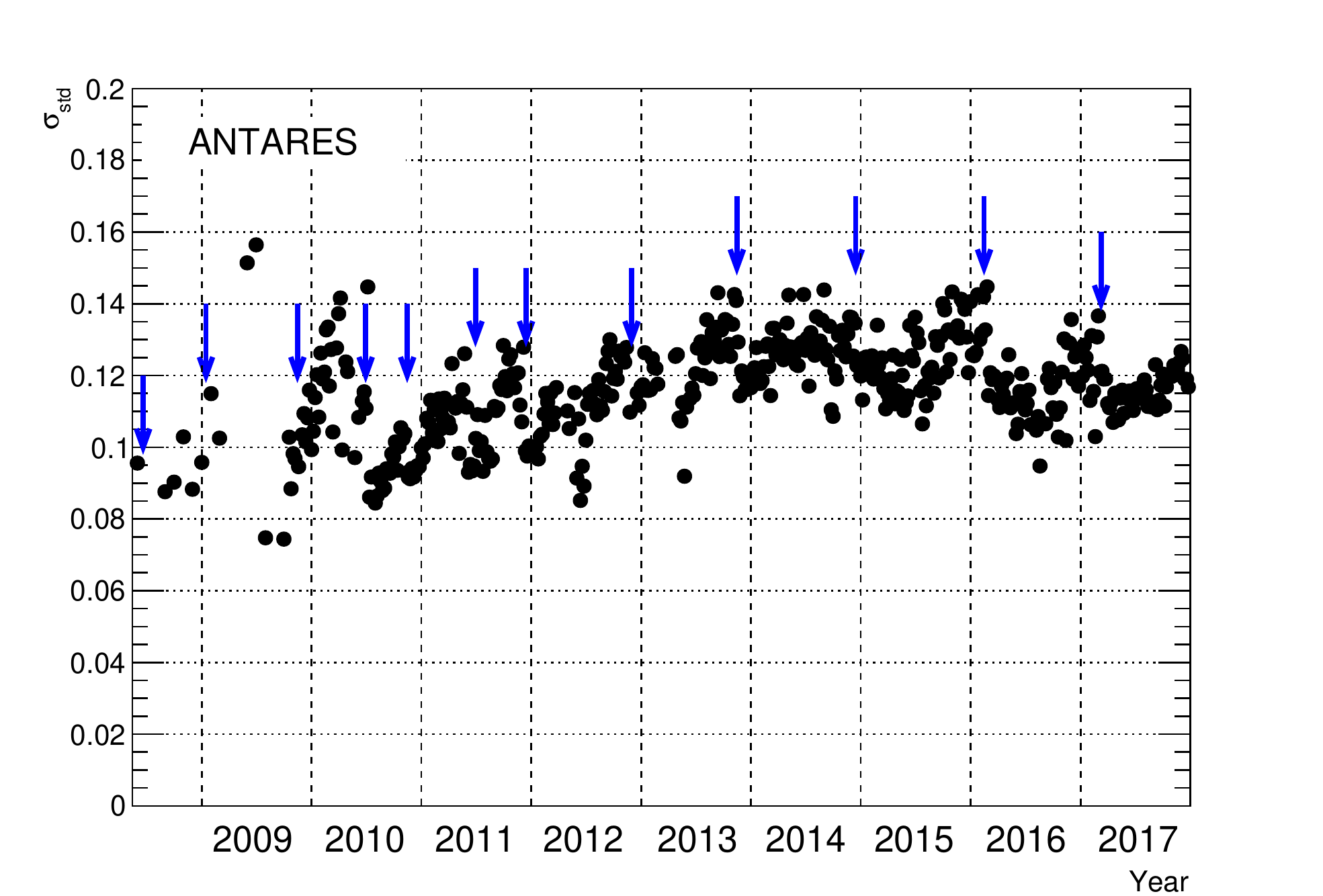}
\caption{Standard deviation $\sigma_{\rm std}$ of the relative photon detection efficiency as a function of time. The blue arrows indicate the periods in which high voltage tuning of the PMTs has been performed.}
\label{fig5}
\end{minipage}
\end{figure}

When averaging the efficiencies of individual PMTs over the whole detector, two statistical quantities are considered: the standard deviation and the error on the mean. The standard deviation is given by:

\begin{equation}
\sigma_{\rm{std}} = \sqrt{\frac{1}{N-1}\sum_i (\epsilon_{i} - \overline{\epsilon})^2}, 
\end{equation} 

\noindent where $N$ is the total number of considered OMs for a given period and $\epsilon_{i}$ is the efficiency of OM $i$ in that particular period. Thanks to the HVT procedure, the $\sigma_{\rm{std}}$ remains stable at the order of 10\%, as can be seen in Figure~\ref{fig5}, where this quantity is shown for all analysed periods. This justifies the fact that we used the average efficiency of the OMs on a given line for those working OMs whose efficiency cannot be computed through the $^{40}\mathrm{K}$ calibration.

\newpage

The statistical error shown in Figure~\ref{fig4} is the error on the mean, defined as:

\begin{equation}
\sigma_{\rm mean} = \frac{\sigma_{\rm std}}{\sqrt{N}}.
\label{eq:Eq11}
\end{equation} 

\noindent This quantity is typically of the order of 1\%.

Possible systematic uncertainties on the individual OM efficiencies, based on the assumption on the Gaussian shape for the distribution of the coincidence rates, have been considered as well. In addition to the intrinsic width of the coincidence peak due to the arrival of photons from a $^{40}\mathrm{K}$ decay process, the shape of the distribution is also affected by photon scattering and the time response of the PMTs. In order to account for these effects, the analysis considers an additional Gaussian term with a larger width compared to that already shown in Equation~\ref{eq:fit}. Using as genuine coincidence rate the sum of the areas under the two Gaussians, the systematic uncertainty has been evaluated as the difference between the resulting efficiencies from the two procedures. Generally it is found that the area under the leading Gaussian is equal to the one of the one Gaussian fit, and the area under the second Gaussian is compatible with zero within its error.

It is worth mentioning that the effects due to some known artefacts from a typical PMT response~\cite{Hamamatsu}, such as delayed hits, remain undetectable in the narrow time window used in this work. The calculated efficiencies exclude these hits, which are later added in the ANTARES simulation chain.

\section{Time calibration}
\label{sec:4}
In order to meet the target angular resolution of the detector, a time syncronisation between all detector components better than 1\,ns is required~\cite{TimeCal}. To achieve such precision a master clock system, located onshore, provides a common reference time to all the offshore electronics, via a network of optical fibers. Further calibration is needed for the delay between the time when the hit is detected and the photon arrival time at the photocathode of the PMT. These time calibrations were performed during the detector construction and are continuously monitored on a weekly basis~\cite{TimeCal}. Two systems of external light sources are used: LED beacons located on four storeys on each detector line, and laser beacons located on the base, at the bottom of several detection lines. They allow to perform both inter-line and inter-storey time calibrations~\cite{OptBeac}. Reconstructions of tracks from downward going muons created in cosmic-ray interactions in the atmosphere are used as well to determine inter-line and inter-storey time calibrations~\cite{TimeCalb}. The $^{40}\mathrm{K}$ method completes the time calibration chain by providing intra-storey timing. All three methods combined assure a precision level of 0.5\,ns for each individual PMT.

\begin{figure}[h!]
\centering
\begin{minipage}[b]{1.\linewidth}
\includegraphics[width=\linewidth]
{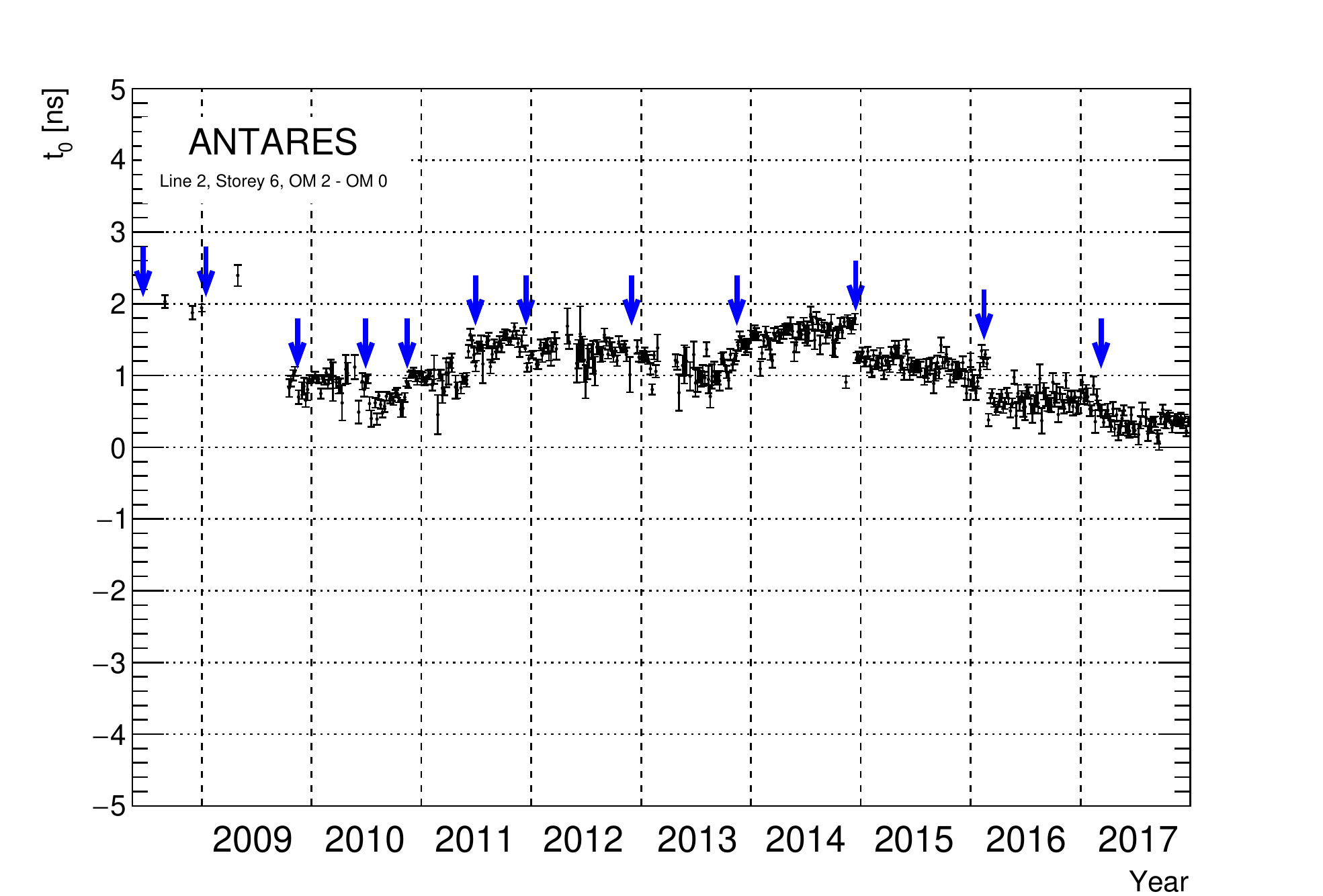}
\caption{Fitted time offset as a function of time for one OM pair (Line 2, Storey 6, OM 2 - OM 0). The blue arrows indicate the periods in which high voltage tuning of the PMTs has been performed.}
\label{fig6}
\end{minipage}
\end{figure}

In Figure~\ref{fig6} the distribution of the fitted time offset, $t_0$, obtained from the $^{40}\mathrm{K}$ coincidence histogram of one OM pair is shown as a function of time.

It can be seen that between subsequent HVT procedures the time difference between two adjacent OMs is stable and its value can be monitored to better than 0.5\,ns. These values, which are produced for each OM pair, serve as input for the intra-storey calibration procedure.

\begin{figure}[h!]
\centering
\begin{minipage}[b]{1.\linewidth}
\includegraphics[width=\linewidth]
{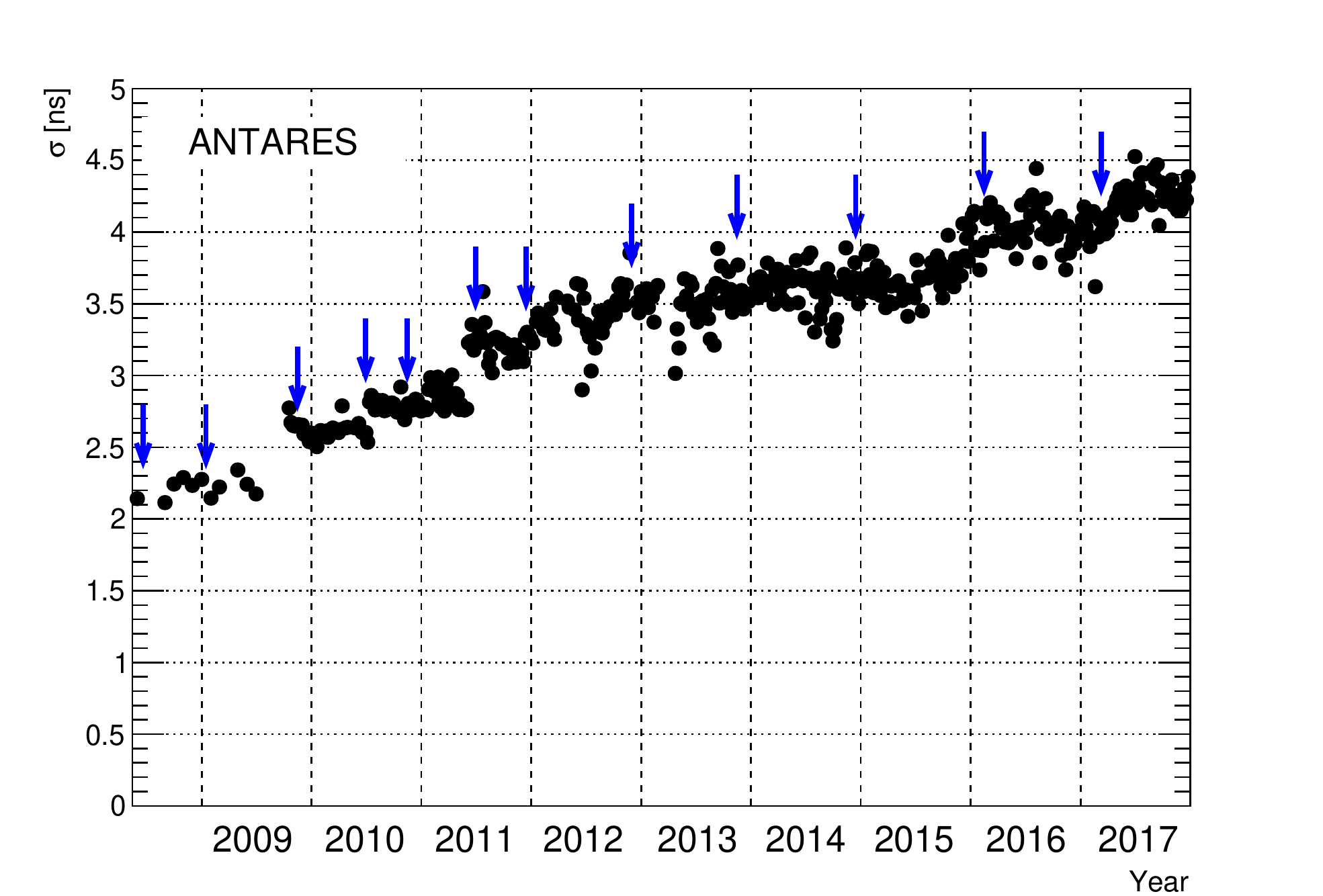}
\end{minipage}
\caption{Standard deviation, $\sigma$, of the average time offset as a function of time. The blue arrows indicate the periods in which high voltage tuning of the PMTs has been performed.}
\label{fig7}
\end{figure}

The standard deviation of the time offset distribution, averaged over the whole detector, is illustrated in Figure~\ref{fig7}. The apparent trend is an increase of the standard deviation as a function of time (from $\sim$ 2 to $\sim$ 4.5\,ns in 9 years). This increase could be correlated to the HVT procedure, which is performed in order to keep the OM detection efficiency at their best. This procedure, in fact, adjusts each individual PMT gain by affecting their transit time, resulting in an overall increase of the standard deviation.

\section{Conclusions and outlook}
\label{sec:5}
Using data collected by the ANTARES neutrino telescope with a dedicated $^{40}\mathrm{K}$ trigger, the photon detection efficiencies for all OMs have been computed from mid 2008 to the end of 2017. The paper presents the stability of a PMT based detector in the hostile environment of the deep-sea, for the longest period ever recorded. It demonstrates that future underwater experiments can remain in operation for timescale of at least a decade without major efficiency degradation. An average decrease of the OM efficiency by $20\%$, as observed from 2008 to 2017, implies a loss of only 15\% in the detection efficiency of an astrophysical signal with a full sky $E^{-2}$ spectrum. The effect of PMT ageing is surely present. The best way to test the biofouling development is the recovery the the OMs, which is planned at the end of the ANTARES physical operation, around 2020.

The results of this study serve as input for detailed Monte Carlo simulations of the ANTARES detector, which include a realistic simulation of the OM efficiencies in each data taking run. The $^{40}\mathrm{K}$ method is also part of the time calibration of the detector.

This procedure can be also exploited in KM3NeT, the next-generation neutrino telescope in the Mediterranean Sea~\cite{LoI}. KM3NeT will consist of two main detectors, ARCA in Sicily, devoted to high energy astroparticle physics, and ORCA in France, focused on few-GeV atmospheric neutrino studies. They both use a configuration similar to the one of ANTARES, but with 31 instead of three PMTs on each storey. This will allow to collect not only double coincidences from $^{40}\mathrm{K}$ decays, but also higher multiplicities, improving the technique to determine the photon detection efficiencies.

\bibliographystyle{unsrt}
\bibliography{./Bibliography}

\end{document}